\documentstyle[aps,prl,epsfig,twocolumn]{revtex}

\def\be{\begin{equation}}
\def\ee{\end{equation}}
\def\bea{\begin{eqnarray}}
\def\eea{\end{eqnarray}}
\def\bi{\begin{itemize}}
\def\ei{\end{itemize}}
\def\bc{\begin{center}}
\def\ec{\end{center}}
\def\bma{\begin{mathletters}}
\def\ema{\end{mathletters}}
\def\one{{I\hspace{-2.1mm}I}}   %identity
\def\C{\hbox{$\mit I$\kern-.6em$\mit C$}}

\begin{document}
%\draft

\title{Separability in 2$\times N$ composite quantum systems}

\author{B. Kraus$^1$, J. I. Cirac$^1$, S. Karnas$^2$, and M.
Lewenstein$^2$}

\address{$^1$Institut f\"ur Theoretische Physik, Universit\"at
Innsbruck, A-6020 Innsbruck, Austria}
\address{$^2$Institut f\"ur Theoretische Physik, Universit\"at Hannover,
D-30163 Hannover, Germany}

\date{\today}

\maketitle

\begin{abstract}
We analyze the separability properties of density operators
supported on $\C^2\otimes \C^N$ whose partial transposes are
positive operators. We show that if the rank of $\rho$ equals
$N$ then it is separable, and that bound entangled states have
rank larger than $N$. We also give a separability criterion for
a generic density operator such that the sum of its rank and the
one of its partial transpose does not exceed $3N$. If it exceeds
this number we show that one can subtract product vectors until
decreasing it to $3N$, while keeping the positivity of $\rho$
and its partial transpose. This automatically gives us a
sufficient criterion for separability for general density
operators. We also prove that all density operators that remain
invariant after partial transposition with respect to the first
system are separable.
\end{abstract}

%\narrowtext

% -------------------------------------------------------------

\section{Introduction}

Perhaps, entanglement is the most intriguing property of Quantum
Mechanics. It arises when two or more systems are in a non--separable
state; that is, in a state that cannot be prepared locally \cite{We89}.
Apart from having fundamental implications\cite{Peresbook}, such as
Bell's theorem or the existence of decoherence, entanglement is in the
realm of several practical applications of Quantum Information
\cite{Di95}.

Despite of its importance, we do not know yet how to quantify
entanglement. In fact, even in the simple case in which we only
have two systems $A$ and $B$ in a mixed state $\rho$, there
exists no general criterion that allows us to distinguish
whether the state is separable or not; that is, whether it can
be written as a mixture of product vectors of the form
$|e,f\rangle\equiv |e\rangle_A\otimes|f\rangle_B$. Several
necessary conditions for separability have been formulated in
the recent years. Werner \cite{We89} formulated a criterion
based on the analysis of a local hidden variable model and on
the positivity of the mean value of the flipping operator. The
Horodeckis \cite{Ho96a} obtained a criterium in the form of
inequalities for the so called $\alpha$-entropies. An important
step forward to clarify the situation was taken by Peres
\cite{Pe96}. Peres criterion is based on the positivity of the
operator obtained after partially transposing the density
operator. This criterion gives a {\em necessary condition} for
separability for density operators in Hilbert spaces of
arbitrary dimension: if the partial transpose of a given density
operator $\rho$ is not positive, then $\rho$ is not separable.

The Horodecki family \cite{Ho96} has established the connection
between separability and the theory of positive operator maps
\cite{St63}. A necessary and sufficient condition for
separability has been expressed in terms of these maps; such
condition, unfortunately, is constructive only if the structure
of the set of positive maps is sufficiently known. This turns
out to be the case for the situations in which the Hilbert space
corresponding to the first subsystem is two--dimensional and the
one of the second is either two-- or three-- dimensional, the
so--called $2\times 2$ (two qubits) and $2\times 3$ cases (one
qubit and one qutrit), respectively. Using that result, the
Horodeckis have shown that Peres criterion (called hereafter
Peres--Horodecki criterion) provides a sufficient condition for
separability in $2\times 2$ and $2\times 3$ systems. The fact
that this is not a sufficient condition for higher dimensional
systems was pointed out later on \cite{Ho98b}, by giving
examples of density matrices in $2\times 4$ and $3\times 3$
spaces that are entangled, but have positive partial transposes
(PPT). After that, a systematic method of constructing entangled
states with PPT, based on the existence of unextendible basis of
product vectors was formulated \cite{Be98,Ho98}. Recently,
positive map theory has been used to obtain a reduction
criterion of separability \cite{Ho99a}.

So far, there have been very few results concerning {\em
sufficient conditions} for separability in high dimensional
systems \cite{Zy98,Ho99}. In this paper we analyze the
separability properties of density operators acting on
$\C^2\otimes \C^N$. This situation is particularly useful since
it allows to determine a number of separability properties in
the multiqubit case \cite{Du99}. Our analysis is based on
subtracting product vectors from $\rho$ while keeping $\rho$ and
its partial transpose positive. The idea of subtracting product
vectors was developed in \cite{Le98} and has proven to be very
useful so far since it allowed: i) to define unique optimal
decomposition of mixed states in Hilbert space of arbitrary
dimensions into a sum of separable and inseparable part; ii) to
find minimal decompositions into product vectors of separable
density operators acting on $\C^2\otimes \C^2$ (two qubits), and
iii) to define a measure of entanglement in those systems. In
the present paper, by studying the conditions under which we can
subtract such product vectors we are able to find necessary and
sufficient conditions for separability for low and medium rank
density operators. We also give a sufficient condition for
separability for general density operators. Furthermore, we show
that when quantum systems remain invariant after partial
transposition with respect to the first system then they are
separable. This last result displays the difference between the
$2\times N$ case and the general $M\times N$ one ($N,M>2$),
where it has been found that there are operators satisfying this
condition which are not separable \cite{Be98}.

The structure of this paper is the following: In Section II we
give the definitions that we use throughout the paper. In
Section III we derive a series of lemmas concerning product
vectors. First, we recover the results given in Ref.\
\cite{Le98} concerning the conditions under which one can
subtract product vectors from a density operator. We then give a
separability criterion for the case in which there is a finite
set of product vectors that can be subtracted. Next we show that
if there are product vectors in the kernel of $\rho$, then the
problem of separability can be highly simplified. Finally, we
analyze under which conditions there exist product vectors in
subspaces (e.g. the kernel and the range of $\rho$), and show
that the problem reduces to finding the roots of complex
polynomials. Section IV contains most of the results of this
work concerning separability. We first show that one just have
to consider the cases in which the rank of $\rho$ is larger or
equal than $N$. When the rank of $\rho$ is equal to $N$ (and
$\rho$ is not trivially supported on a smaller subsystem of $B$)
then it is separable. From this result follows that bound
entangled states have a rank larger than $N$. We continue by
analyzing the case in which the sum of the ranks of $\rho$ and
its partial transpose is smaller than $3N$, giving a necessary
and sufficient separability criterion for generic density
operators and for $N\le 10$. After that, we show how in the
general case one can reduce the problem to that situation. Then,
we prove that if $\rho$ is equal to its partial transpose with
respect to the first subsystem, then it is separable. The same
statement holds when $\rho$ is ``sufficiently close" to its
partial transpose (in the sense of operator norm, for instance).
Finally, in Section V we illustrate our results by analyzing the
case $\C^2\otimes\C^4$.

% =========================================================
\section{Separability in $2\times N$}

We consider a density operator $\rho \geq 0$ acting on $\C^2
\otimes \C^{N}$. We wish to find out conditions such that
we can state whether $\rho$ is separable or not; that is,
whether it can be written as
\be
\label{rhosep}
\rho=\sum_i |e_i,f_i\rangle\langle e_i,f_i|,
\ee
where $|e_i,f_i\rangle\equiv |e_i\rangle_A\otimes|f_i\rangle_B$ are
unnormalized product states. Throughout this paper we will make use of
the fact that a necessary condition for separability is that the partial
transposition of $\rho$ is a positive operator \cite{Pe96}. Given an
orthonormal basis $\{|0\rangle,|1\rangle\}\in \C^2 $, we define the
partial transposition of $\rho$ with respect to the first system in that
basis as
\be
\rho^{T_A} =\sum_{i,j=0}^1 \phantom{i}_{A}
\langle i| \rho| j\rangle_{A} |j\rangle_{A}\langle i|
\ee
Thus, from now on we will assume that both $\rho,\rho^{T_A}\ge
0$ (otherwise $\rho$ is already not separable).

Throughout this paper $K(X), R(X), k(X)$, and $r(X)$ denote the kernel,
range, dimension of the kernel, and rank of the operator $X$,
respectively. By $| e^{*}\rangle$ we will denote the complex conjugated
vector of $| e\rangle$ in the basis $|0\rangle_{A},|1\rangle_{A}$ in
which we perform the partial transposition; that is, if $| e\rangle
=\alpha| 0\rangle+\beta | 1\rangle$ then $| e^{*}\rangle =\alpha^{*}| 0
\rangle+\beta^{*} | 1\rangle$. The subscript "$r$" will be used to
denote real vectors; that is $| e_{r}\rangle =|e^{*}_{r}\rangle$. We will
denote by $|\hat{e} \rangle$ the vector which is orthogonal to $|e\rangle$.
On the other hand, we will make use of the following property,
\be
\label{eq:partialt}
\phantom{i}_{A}\langle e_{i} | \rho| e_{j}\rangle_{A}=
\phantom{i}_{A}\langle e^{*}_{j} | \rho^{T_{A}}| e^{*}_{i}\rangle_{A}
\ee
for any pair $| e_{i}\rangle, | e_{j}\rangle \in \C^2$, which can be
easily deduced from the definition of partial transposition. Note that
if $\rho$ is separable [cf (\ref{rhosep})], then
\be
\rho^{T_A}=\sum_i |e_i^\ast,f_i\rangle\langle e_i^\ast,f_i|,
\ee
is also separable; obviously, the opposite is also true.

We say that $\rho$ acting on $\C^2\otimes\C^N$ is supported in
$\C^2\otimes\C^M$ if the minimal subspace $H\subseteq \C^N$ such
that $R(\rho)\subseteq \C^2\otimes H$ has dimension $M$. This
means that there exist $N-M$ (and not more) linearly independent
vectors $\{|f_i\rangle,i=1,2,\ldots,N-M\}\in \C^N$ such that
$|e,f_i\rangle\in K(\rho)$ for all $|e\rangle\in\C^2$. The
utility of this definition can be seen in the following example.
Let us consider an operator $\rho$ acting on $\C^2\otimes\C^2$.
We can always trivially regard this operator as acting on a
higher dimensional space (e.g., $\C^2\otimes\C^4$) but giving
zero whenever it acts outside the original space. If we are
studying the separability properties of $\rho$ it is useful to
get rid of the spurious dimensions and really consider $\rho$ as
acting on $\C^2\otimes\C^2$. Thus, we will concentrate on
operators not only acting, but also supported on $\C^2\otimes
\C^N$.

The method we use to find separability conditions is based on
subtracting product vectors from $\rho$ in such a way that both $\rho$
and $\rho^{T_A}$ remain positive. If we are able to subtract several of
those product vectors until we achieve that the result vanishes, then we
will have obtained that $\rho$ is separable. Thus, a crucial point is to
find the conditions under which we can find such product vectors. In the
next few sections we will determine those conditions.

% =======================================================
\section{Product Vectors}

As mentioned above, our procedure to study separability is based on
subtracting product vectors $|e,f\rangle$ from the density operator
$\rho$ in such a way that the resulting operator and its partial
transpose remain positive. Thus, we have to determine under which
conditions this is possible. This section deals with that problem. We
have divided it in three parts: (i) Properties related to the existence
of product vectors in the range of $\rho$ or $\rho^{T_A}$; (ii)
Properties related to the existence of product vectors in the kernel of
$\rho$ or $\rho^{T_A}$; (iii) Existence of product vectors in subspaces.
In the first subsection we derive some useful properties of a density
operator whenever there are product vectors in its range. In particular,
we will recover the results of Ref.\ \cite{Le98} which state that we can
only subtract product vectors of $\rho$ that belong to its range, and
that by subtracting the right amount we can always decrease its rank.
Then we will show that if we have a set of product vectors in the range
of $\rho$ and another related set in the range of $\rho^{T_A}$, then
$\rho$ is separable iff it can be written as a mixture of those vectors.
This means that if we can find all the product vectors satisfying these
conditions, then we can check whether $\rho$ is separable or not by
using only those vectors. In the second subsection we will derive some
other properties of $\rho$ whenever there are product vectors in its
kernel. In particular we will show that in that case one can subtract
one product vector and not only reduce the rank of $\rho$, but also
reduce the problem to finding whether a related density operator
supported on $\C^2\times \C^{N-1}$ is separable. This means that the
product vectors in the kernel play a very important role since they
allow to simplify the problem of separability. Finally, in the last
subsection we will derive the conditions needed to have product vectors
in subspaces; in particular in the kernel and the range of $\rho$ and
$\rho^{T_A}$. We will see that in order to find them, one has to solve
some complex polynomial equations. The solutions to those equations
determine which are the product vectors that can be subtracted from
$\rho$, which, together with the results of the first subsection, allow
us to analyze the separability of certain density operators, as it will
be shown in the next section.

% ========================================================
\subsection{Range of $\rho$}

We have divided this subsection in two: In the first one we introduce a
lemma which gives the conditions under which the difference of two
positive operators is a positive operator itself. This allows us to
recover in a simple way the results given in \cite{Le98} regarding
subtracting product vectors from $\rho$. In the second one we show that
the set of (common) product vectors in the ranges of $\rho$ and
$\rho^{T_A}$ suffices to study the separability properties of $\rho$.

% ========================================================
\subsubsection{Subtracting product vectors}

We use the operator norm, namely $\Vert A \Vert=\mbox{max}_{\Vert |
\phi\rangle \Vert=1}\Vert A| \phi\rangle \Vert$. For a given hermitian
operator $X$ we will denote by $X^{-1}$ its pseudoinverse; that is, if
the spectral decomposition of $X=\sum_k X_k |x_k\rangle\langle x_k|$
with $X_k\ne 0$, then $X^{-1}=\sum_k X_k^{-1} |x_k\rangle\langle x_k|$.

{\bf Lemma 1:} Given two hermitian operators $X,Y\geq 0$ then $X-Y\geq
0$ iff $K(X)\subseteq K(Y)$ and $\Vert Y^{\frac{1}{2}}
X^{-\frac{1}{2}}\Vert^{2} \leq 1$.

{\em Proof:} (if) We can write an arbitrary vector
$|\Psi\rangle=|\Psi^r\rangle + |\Psi^k\rangle$, where $|\Psi^r\rangle\in
R(X)$ and $|\Psi^k\rangle\in K(X)\subseteq K(Y)$. Therefore $\langle
\Psi|X-Y|\Psi\rangle = \langle \Psi^r|X-Y|\Psi^r\rangle$. Since
$|\Psi^r\rangle\in R(X)=R(X^{1/2})$ we have that there exists
some $|\Phi\rangle \in R(X^{1/2})$ such that
$|\Psi^r\rangle=X^{-1/2}|\Phi\rangle$. Using that
\be
\langle \Phi| X^{-1/2}Y X^{-1/2}|\Phi\rangle \le \langle
\Phi|\Phi\rangle,
\ee
we obtain that $\langle \Psi^r|X-Y|\Psi^r\rangle\ge 0$, and thus $X-Y\ge
0$. (only if) First, $K(X)\subseteq K(Y)$ since otherwise there would be
a vector $|\Psi\rangle\in K(X)$ for which $\langle \Psi|Y|\Psi\rangle
>0$, and therefore $\langle \Psi|X-Y|\Psi\rangle <0$. Moreover, $\forall
|\Psi^r\rangle \in R(X)$ we can find $|\Phi\rangle \in
R(X^{1/2})$ such that $|\Psi^r\rangle=X^{1/2}|\Phi\rangle$.
Using that $\langle
\Phi|X|\Phi\rangle \ge \langle \Phi|Y|\Phi\rangle$ we obtain that
\be
1\ge \langle\Psi^r|X^{-1/2}YX^{-1/2}
|\Psi^r\rangle/\langle\Psi^r|\Psi^r\rangle,
\ee
which immediately gives $\Vert Y^{\frac{1}{2}}X^{-\frac{1}{2}}\Vert^{2}
\leq 1$.

We can use this lemma to recover the results of Ref.\ \cite{Le98}. Given
an operator $\rho$, a product vector $|e,f\rangle$ and $\lambda>0$, we
define
\bma
\bea
\label{subs}
\tilde \rho &\equiv&
\rho-\lambda|e,f\rangle\langle e,f|,\\
\lambda_0 &=& \frac{1}{\langle e,f| \rho^{-1}| e,f \rangle},\\
\bar\lambda_0 &=& \frac{1}{\langle e^\ast,f| (\rho^{T_A})^{-1}|e^\ast,f
\rangle}.
\eea
\ema
We then have

{\bf Corollary 1:} $\tilde\rho,\tilde\rho^{T_A}\ge 0$ iff
$|e,f\rangle\in R(\rho)$, $| e^{*},f\rangle \in R(\rho^{T_{A}})$ and
$\lambda \le \mbox{min}(\lambda_0,\bar\lambda_0)$.

{\em Proof}: We can simply use Lemma 1 with $X=\rho$ and $Y=\lambda|
e,f\rangle\langle e,f|$, and with $X=\rho^{T_A}$ and $Y=\lambda|
e^\ast,f\rangle\langle e^\ast,f|$.$\Box$

The most interesting situation occurs when we choose
$\lambda=\mbox{min}(\lambda_0,\bar\lambda_0)$, since in that case we can
reduce the ranks of $\rho$ and/or $\rho^{T_A}$, as states the following

{\bf Lemma 2:} Let $|e,f\rangle\in R(\rho)$, $| e^{*},f\rangle \in
R(\rho^{T_{A}})$. Then

\begin{description}

\item[(i)] If $\lambda=\lambda_0<\bar\lambda_0$ then
$r(\tilde\rho)=r(\rho)-1$ and $r(\tilde\rho^{T_A})=r(\rho^{T_A})$.

\item[(ii)] If $\lambda=\bar\lambda_0<\lambda_0$ then
$r(\tilde\rho)=r(\rho)$ and $r(\tilde\rho^{T_A})=r(\rho^{T_A})-1$.

\item[(iii)] If $\lambda=\bar\lambda_0=\lambda_0$ then
$r(\tilde\rho)=r(\rho)-1$ and $r(\tilde\rho^{T_A})=r(\rho^{T_A})-1$.

\end{description}

{\em Proof:} Here, we just prove (i), since for the other two
cases one can follow the same arguments. First, note that
$r(\tilde\rho)\ge r(\rho)-1$ since they differ only by one
one--dimensional projector. On the other hand, $K(\rho)\subseteq
K(\tilde\rho)$ for if $|\Psi\rangle\in K(\rho)$ then
$\tilde\rho|\Psi\rangle=-\lambda_0\langle e,f|\Psi\rangle=0$
since $|e,f\rangle\in R(\rho)\perp K(\rho)$. Besides, the vector
$\rho^{-1}|e,f\rangle$ is not in $K(\rho)$ but it is indeed in
$K(\tilde\rho)$, which can be checked by substitution in Eq.\
(\ref{subs}). $\Box$

This lemma allows us to decrease either the rank of $\rho$, of its
partial transpose, or of both of them by subtracting a product vector
$|e,f\rangle$ which belongs to $R(\rho)$ and such that
$|e^\ast,f\rangle$ is in $R(\rho^{T_A})$.

% ========================================================
\subsubsection{Separability and product vectors}

Let us denote by $\{| e_i, f_i\rangle, i=1,2,\ldots,L\}$ the product
vectors such that $| e_i, f_i\rangle\in R(\rho)$ and $| e_i^\ast,
f_i\rangle\in R(\rho^{T_A})$. Note that we assume here that the total
number of such vectors $L$ is finite. We also denote by $ P_i\equiv|
e_i, f_i\rangle\langle e_i, f_i|$. We have

{\bf Lemma 3:} $\rho$ is separable iff it can be written as a convex
combination of $ P_i$.

{\em Proof:} (if) Use the definition of separable state; (only if) We
just have to notice that if a positive operator $\rho$ can be written as
a convex combination of certain projectors on the vectors
$|\Psi_i\rangle$, then these vectors must belong to its range.
Otherwise, there would exist a certain vector
$|\Psi_{i_0}\rangle=|\Psi_{i_0}^r\rangle+|\Psi_{i_0}^k\rangle$, where
$|\Psi_{i_0}^r\rangle\in R(\rho)$ and $0\ne |\Psi_{i_0}^k\rangle\in
K(\rho)$, so that $\langle \Psi_{i_0}^k|\rho|\Psi_{i_0}^k\rangle >0$,
which contradicts the fact that $|\Psi_{i_0}^k\rangle\in K(\rho)$. Thus,
if $\rho$ is separable then it can be written as a convex combination of
projectors on product vectors $|e_i,f_i\rangle \in R(\rho)$. Also
$\rho^{T_A}$ can be written as the same convex combination, but with
$|e_i^\ast,f_i\rangle\in R(\rho^{T_A})$.$\Box$

Let the projectors $P_i$ be linearly independent. These
projectors belong to the space ${\cal L}[R(\rho)]$ of linear
operators acting on $R(\rho)$, which has dimension $r(\rho)^2$.
On the other hand, the projectors $P_i^{T_A}$ will be linearly
independent as well, since partial transposition is a linear
invertible map. These projectors belong to the space ${\cal
L}[R(\rho^{T_A})]$, which has dimension $r(\rho^{T_A})^2$.
Therefore, if the $P_i$ are linearly independent we must have
$L\le {\rm min} [r(\rho)^2,r(\rho^{T_A})^2]$. We can always
complete the set $P_i$ with other operators $P_i$
($i=L+1,\ldots,r(\rho)^2$) to form a basis in ${\cal
L}[R(\rho)]$. That is a Hilbert space with the scalar product
$(A,B)={\rm tr}(A^\dagger B)$. Thus, we can construct the
biorthogonal basis to $\{P_i\}$ in ${\cal L}[R(\rho)]$; that is,
we can find a set of operators $Q_i\in {\cal L}[R(\rho)]$ such
that tr$(Q_i^\dagger P_j)=\delta_{ij}$. Then we have the
following necessary and sufficient separability condition

{\bf Lemma 4:} If $P_i$ are linearly independent, then $\rho$ is
separable iff tr$(Q_i^\dagger \rho)\ge 0$ $\forall i\le L$ and
tr$(Q_i^\dagger \rho)= 0$ $\forall i> L$.

{\rm Proof:} Since the $P_i$ form a basis, $\rho$ can be
expanded in a unique way as
\be
\label{rhorho}
\rho = \sum_{i=1}^{r(\rho)^2} c_i P_i,
\ee
with $c_i={\rm tr}(Q_i\rho)$. (if) since $c_i\ge 0$ we have that $\rho$
can be written as a convex combination of projectors on $| e_i,
f_i\rangle$, i.e. it is separable; (only if) If $\rho$ is separable,
according to Lemma 3 it can be written as (\ref{rhorho}) with $c_i=0$ if
$i>L$ and $c_i\ge 0$ if $i\le L$.$\Box$

This lemma is important since in general, if $L\le {\rm
min}[r(\rho)^2,r(\rho^{T_A})]$, the projectors will be linearly
independent. The reason is that given a set of $L\le D^2$ random
product vectors which span a Hilbert space of dimension $D$,
then the corresponding projectors are linearly independent.

%================================================
\subsection{Kernel of $\rho$}

We will show that if there is a product vector in the kernel of $\rho$,
then one can find a positive operator acting on $\C^2\otimes \C^{N-1}$
which has the same separability properties as $\rho$. This means that
one can reduce the dimensionality of the Hilbert space of the second
subsystem, which simplifies the problem. We will use later on these
results to prove facts regarding low rank density operators. We start
proving a simple

{\bf Lemma 5:} $|e^\ast,f \rangle \in K(\rho^{T_{A}})$ iff $|e,f\rangle
\in K(\rho)$.

{\em Proof:} (if) We have $0= \langle e,f | \rho | e,f \rangle$. Using
(\ref{eq:partialt}), we have $0=\langle e^{*},f |\rho^{T_{A}} | e^{*},f
\rangle$ and since $\rho^{T_A}\ge 0$ then $\rho^{T_{A}}|
e^{*},f\rangle=0$; (only if) One can prove it in the same way.$\Box$

{\bf Lemma 6:} If $| e,f \rangle \in K(\rho)$ then either:

\begin{description}

\item[(i)] $| \hat{e},f \rangle \in K(\rho)$, or

\item[(ii)] there exists a non-zero $| g \rangle \in \C^{N} $ such that $\rho |
\hat{e},f\rangle = | \hat{e},g\rangle$ and $\rho^{T_{A}} |
\hat{e}^{*},f\rangle = | \hat{e}^{*},g\rangle$.

\end{description}

{\em Proof:} According to Lemma 5, if $| e,f \rangle \in
K(\rho)$ then $|e^{*},f\rangle \in K(\rho^{T_{A}})$. Therefore
\be
0=\langle \hat{e}^{*},h | \rho^{T_{A}} | e^{*},f
\rangle=\langle e,h | \rho | \hat{e},f \rangle,
\ee
$\forall | h\rangle \in \C^{N}$. Thus we know that either $\rho
| \hat{e},f \rangle=0$ or $\rho | \hat{e},f \rangle=| \hat{e},g_{1}
\rangle $ for some $| g_{1} \rangle \in \C^{N} $. Analogously, since
$|e,f \rangle \in K(\rho)$ we have that either $\rho^{T_{A}}|
\hat{e}^{*},f \rangle=0$ or $ \rho^{T_{A}} |
\hat{e}^{*},f \rangle=| \hat{e}^{*},g_{2} \rangle $ for some $|
g_{2}\rangle \in \C^{N}$. In the first case, using Lemma 5 we
have (i). For (ii), it remains to be proven that $|g_{1}\rangle
=| g_{2} \rangle$. Using the orthogonality of $| e \rangle$ and $|
\hat{e} \rangle$ we have
\be
\Vert | \hat{e}\rangle \Vert^{2} | g_{1} \rangle
=\langle \hat{e} | \rho | \hat{e},f \rangle =\langle
\hat{e}^{*}| \rho^{T_{A}} | \hat{e}^{*},f \rangle=
\Vert | \hat{e}\rangle \Vert^{2} | g_{2} \rangle.\Box
\ee

The first part of this lemma simply states that $\rho$ is
supported on a smaller space $\C^2\otimes\C^{M}$ with $M<N$,
since $|e,f\rangle\in K(\rho)$ for all $|e\rangle\in\C^2$. In
other words, that we have spurious dimensions in the second
subsystem. In the following, we will not consider that trivial
case, and therefore we will assume that $\rho$ is supported on
the whole space $\C^2\times\C^N$. In that case, Lemma 6 together
with Lemma 2 tell us that there is a vector that can be
subtracted from $\rho$. In fact, this gives rise to one of the
most important results of this section:

{\bf Lemma 7:} If $\rho$ is supported in $\C^2\otimes\C^N$ and there
exists a product vector in $K(\rho)$ then $\rho=\tilde\rho+\rho_{s}$,
where $\rho_{s}$ is a projector on a product vector and

\begin{description}
\item[(i)] $r(\tilde\rho)=r(\rho)-1$ and $r(\tilde\rho^{T_A})=r(\rho^{T_A})-1$.
\item[(ii)] $\tilde\rho$ is supported on $\C^2 \otimes\C^{N-1}$.
\item[(iii)] $\tilde\rho$ is separable iff $\rho$ is separable.
\end{description}

{\em Proof:} We have that $|e,f\rangle\in K(\rho)$. Using the previous
Lemma (ii) (since $\rho$ is supported in $\C^2\otimes\C^N$) we have that
$\rho |\hat{e},f\rangle =| \hat{e},g\rangle$ and $\rho^{T_{A}} |
\hat{e}^{*},f\rangle = | \hat{e}^{*},g\rangle$. According to Corollary 1
we can subtract the product vector $| \hat{e},g\rangle$,
obtaining $\tilde\rho=\rho-\rho_s$ where $\rho_s=\lambda |
\hat{e},g\rangle\langle\hat{e},g|$ and
\bea
\label{UI}
\lambda &=& \lambda_{0}=\frac{1}{\langle \hat{e},g| \rho^{-1}
| \hat{e},g\rangle} = \frac{1}{\Vert |
\hat{e}\rangle \Vert^{2}\langle g | f\rangle}\nonumber\\
&=&\frac{1}{\langle
\hat{e}^{*},g| (\rho^{T_{A}})^{-1} |
\hat{e}^{*},g\rangle}=\bar\lambda_0.
\eea
According to Lemma 2 (iii) we have decreased the ranks of both
$\rho$ and $\rho^{T_A}$ by one, which proves (i). We also have
that $|\hat e,f\rangle,|e,f\rangle\in K(\tilde\rho)$ and
therefore $\tilde\rho$ is supported in $\C^2\otimes\C^{M}$ with
$M<N$. Now we show that $M=N-1$. If it was smaller, then there
would exist $|h\rangle$ orthogonal to $|f\rangle$ such that
$|e,h\rangle,|\hat e,h\rangle\in K(\tilde\rho)$. In that case we
would have $\rho|e,h\rangle=0$ and $\rho|\hat e,h\rangle= c
|\hat e,g\rangle$ where $c$ is a constant. If we define
$|f'\rangle=c|f\rangle-|h\rangle\ne 0$ we would have that
$|e,f'\rangle,|\hat e,f'\rangle\in K(\rho)$, contrary to our
assumption that $\rho$ is supported on $\C^2\otimes\C^N$, which
proves (ii). It remains to be shown (iii) that $\rho$ is
separable iff $\tilde \rho$ is too: (if) trivial; (only if) If
$\rho$ is separable, and since $\rho|e,f\rangle=0$ we can always
write
\be
\rho = \sum |e_i,\hat f_i\rangle\langle e_i,\hat f_i|
+ |\hat e\rangle\langle \hat e| \otimes \eta,
\ee
where $\langle f|\hat f_i\rangle=0$ and $\eta$ is a positive operator
acting on $\C^N$. If we impose $|\hat e,g\rangle= \rho|\hat e,f\rangle$
we obtain $|g\rangle = \Vert |\hat e\rangle\Vert^2 \eta|f\rangle$, and
therefore $|g\rangle\in R(\eta)$. We can write
\be
\tilde \rho = \sum |e_i,\hat f_i\rangle\langle e_i,\hat f_i|
+ |\hat e\rangle\langle \hat e| \otimes
(\eta-\lambda_0|g\rangle\langle g|),
\ee
so that if we show that the operator $(\eta-\lambda_0|g\rangle\langle
g|)\ge 0$ then we have that $\tilde \rho$ is separable. Using (\ref{UI})
we have that such operator is
\be
\eta - \frac{1}{\Vert |\hat{e}\rangle \Vert^{2}\langle g | f\rangle} |g\rangle\langle
g| =\eta - \frac{1}{\langle g| \eta^{-1}|
g\rangle} |g\rangle\langle g|,
\ee
which is positive according to Lemma 1.$\Box$

This theorem simply says that when there is a product vector in the
kernel or $\rho$, then we can reduce the problem without changing the
separability properties of $\rho$. Later on we will consider states
which are invariant under partial transposition. In that case, we can
particularize this result and obtain:

{\bf Lemma 8:} If in addition to the premises of Lemma 7
$\rho=\rho^{T_A}$ then we can always obtain the results of that lemma
with $\tilde\rho=\tilde\rho^{T_A}$.

{\em Proof:} We know from Lemma 5 that if there is a product vector $|
e,f\rangle \in K(\rho)$ then $| e^{*},f \rangle \in
K(\rho^{T_{A}})=K(\rho)$. With these two vectors we can always construct
$|e_r,f\rangle\in K(\rho)$ where $|e_r\rangle$ is real (in the basis in
which we take the partial transposition). If we proceed with this vector
as in the previous lemma then we obtain the desired result. $\Box$

Finally, we will show a property of the vectors contained in $K(\rho)$
for the case when there is no product vector in $K(\rho)$. We will use
this property later on to simplify the search of product vectors in
$R(\rho)$. Let us denote by $\{|\Psi_{i}^{1}\rangle, i=1\ldots
k(\rho)\}$ and $\{|\Psi_{i}^{2}\rangle, i=1\ldots k(\rho^{T_{A}})\}$ a
basis in $K(\rho)$ and $K(\rho^{T_A})$, respectively. Then we have the
following:

{\bf Lemma 9:} If there exists no product vector in $K(\rho)$ then both
$\{\langle e|\Psi_{i}^{1}\rangle\}$ and $\{\langle
e^\ast|\Psi_{i}^{2}\rangle\}$ are sets of linearly independent vectors
in $\C^N$ for any $|e\rangle \in \C^2$.

{\em Proof:} We start proving that for any $|e\rangle \in \C^2$,
the operator $\langle e|\rho|e\rangle$ is invertible. Otherwise
there would exist $|h\rangle\in\C^N$ such that $0=\langle
h|\langle e| \rho|e\rangle |h\rangle$ and since $\rho\ge 0$ then
$\rho|\hat{e},h\rangle=0$, which contradicts our assumptions. On
the other hand, since $\{|e\rangle,|\hat{e}\rangle\}$ is a basis
in $\C^2$ we can write the vectors in the kernel of $\rho$ as
\be
\label{bar}
|\Psi_{i}^{1}\rangle
=|e,\Psi^{e}_{i}\rangle+|\hat{e},\Psi^{\hat{e}}_{i}\rangle.
\ee
Using that they are in the kernel of $\rho$ we find that
\be
|\Psi^{\hat{e}}_{i}\rangle =-\frac{1}{\langle
\hat{e}|\rho|\hat{e}\rangle }\langle \hat{e}|\rho|e\rangle
|\Psi^{e}_{i}\rangle,
\ee
where we made use of the fact that the operator $\langle
\hat{e}|\rho|\hat{e}\rangle $ is invertible. Substituting this
expression in (\ref{bar}), we have
\be
|\Psi_{i}^{1}\rangle =(\one +|\hat{e}\rangle
\frac{1}{\langle
\hat{e}|\rho|\hat{e}\rangle }\langle \hat{e}|\rho)|e,
\Psi^{e}_{i}\rangle.
\ee
Since the $|\Psi_{i_1}^{1}\rangle$'s are linearly independent, so must
be the $|\Psi_{i_1}^{e}\rangle$'s. The same argumentation holds for the
vectors in the kernel of $\rho^{T_{A}}$. $\Box$

%============================================================
\subsection{In Subspaces}

As we have shown above, it is important to know when one can be sure
that there exists a product vector in either the range or the kernel of
a density operator. In this section we will prove that for sufficiently
large subspaces there always exist product vectors. We will denote by
$\{|0\rangle,|1 \rangle\}$, and $\{|1\rangle,\ldots,| N\rangle\}$
orthonormal basis in $\C^2$ and $\C^N$, respectively.

{\bf Lemma 10:} Any subspace $H \subseteq \C^2 \otimes \C^N$ with
dim$(H)=M>N$ contains an infinite number of product vectors. If $M= N $
it contains at least one product vector.

{\em Proof:} We denote by $\{ | \Psi_{i}\rangle, i=1,\ldots
2N-M\}$ a basis in the orthogonal complement of $H$. We have
\be
| \Psi_{i}\rangle =\sum_{k=1}^{N} \left [A_{i,k}| 0,k\rangle+
B_{i,k}| 1,k\rangle \right ],
\ee
where $A$ and $B$ are $(2N-M) \times N$- matrices. We write the product
vector $|e,f\rangle$ as
\be
\label{ef}
|e,f\rangle =\left( \alpha | 0\rangle_A +|1\rangle_{A}\right)\otimes
\sum_{k} f_{k}| k\rangle_{B}.
\ee
Imposing that $|e,f\rangle$ is orthogonal to the $|\Psi_i\rangle$ for
all $i$ gives $(\alpha A^\ast +B^\ast)\vec f=0$. This can be regarded as
an equation for $\vec f$. When $M>N$ there exist non trivial solutions
for each $\alpha$, i.e. for each $|e\rangle$. For $M=N$ we have non
trivial results provided det$(\alpha A^\ast+B^\ast)=0$; the determinant
is a polynomial of $N$--th degree in $\alpha$, which always has roots.
$\Box$

{\bf Corollary 2:} Any subspace $H \subseteq \C^2 \otimes \C^N$ with
dim$(H)=M>N$ contains an infinite number of product vectors of the form
$|e_{r},f\rangle$, where $|e_{r}\rangle=|e^{*}_{r}\rangle$.

As we have seen in previous subsections, we can subtract vectors from
$\rho$ while keeping the positivity of both $\rho$ and $\rho^{T_A}$
provided we can find a product vector $|e,f\rangle\in R(\rho)$ and such
that $|e^\ast,f\rangle\in R(\rho^{T_A})$. We can use the idea given in
the proof of the previous lemma to find out under which conditions this
is possible. Given any two subspaces $H_1,H_2\in \C^2\otimes\C^N$ with
dim$(H_{1,2})=M_{1,2}$ we denote by $\{ |\Psi_{i_{1,2}}^{1,2}\rangle,
i_{1,2}=1,\ldots 2N-M_{1,2}\}$ a basis in the orthogonal complement of
$H_{1,2}$, respectively. Then we have

{\bf Lemma 11:} (i) If $M_1+M_2>3N$ then there exists an
infinite number of product vectors $|e,f\rangle\in H_1$ such
that $|e^\ast,f\rangle\in H_2$; (ii) If $M_1+M_2\le 3N$ then
there exists a product vector $|e,f\rangle\in H_1$ such that
$|e^\ast,f\rangle\in H_2$ iff we can find an $\alpha$ such that
there are at most $N-1$ linearly independent vectors among the
following vectors: $\{\alpha \langle \Psi_{i_1}^{1}|0\rangle +
\langle \Psi_{i_1}^{1}|1\rangle, \alpha^\ast
\langle \Psi_{i_2}^{2}|0\rangle + \langle \Psi_{i_2}^{2}|1\rangle\}$.

{\em Proof:} We have
\be
| \Psi_{i_{1,2}}^{1,2}\rangle = \sum_{k=1}^{N} \left
[A^{1,2}_{i,k}| 0,k\rangle+ B^{1,2}_{i,k}| 1,k\rangle \right],
\ee
where $A^{1,2}$ and $B^{1,2}$ are $(2N-M_{1,2}) \times N$-
matrices. Writing a product vector $|e,f\rangle$ as in
(\ref{ef}) and imposing that it is orthogonal to
$|\Psi_{i}^{1}\rangle$ and $|e^\ast,f\rangle$ is orthogonal to
$| \Psi_{i}^{2}\rangle$ for all $i$, we arrive at the following
equations:
\bma
\bea
\left[ \alpha (A^{1})^\ast + (B^{1})^\ast \right] \vec f &=& 0,\\
\left[ \alpha^\ast (A^{2})^\ast + (B^{2})^\ast \right] \vec f &=& 0,
\eea
\ema
which can be regarded as $4N-M_1-M_2$ equations for $\vec f$.
For $M_1+M_2>3N$, there are less equations than variables, and
thus there always exists a solution for each $\alpha$, i.e. for
each $|e\rangle$, which proves (i). For $M_1+M_2\le 3N$ there
exists a nontrivial solution iff the matrix
$M(\alpha,\alpha^\ast)$ composed of $\alpha (A^{1})^\ast +
(B^{1})^\ast$ and $\alpha (A^{2})^\ast + (B^{2})^\ast$ has rank
smaller than $N$, which is equivalent to the statement of the
lemma.$\Box$

Let us now analyze the case in which $M_1+M_2=3N$. Here, we have that
the conditions of Lemma 11 are fulfilled iff
det$[M(\alpha,\alpha^\ast)]=0$ for some $\alpha$. This determinant will
be a polynomial $P_{2N-M_1,2N-M_2}(\alpha,\alpha^\ast)$ of degree
$2N-M_1$ in $\alpha$ and of degree $2N-M_2$ in $\alpha^\ast$. Thus, the
existence of product vectors $|e,f\rangle\in H_1$ such that
$|e^\ast,f\rangle\in H_2$ is directly related to the existence of roots
of that polynomial. For $M_1+M_2<3N$ we have to impose that several
determinants containing $N$ rows of $M(\alpha,\alpha^\ast)$ vanish;
again, those conditions are equivalent to imposing that several
polynomials in $\alpha$ and $\alpha^\ast$ have common roots.

Note that for complex polynomials there is no theorem which allows to
know how many (if any) roots exist. For example, the equation
$\alpha^\ast\alpha+1$ has no roots, whereas $\alpha^2-(\alpha^\ast)^2=0$
has infinitely many. However, in a generic case, given a polynomial
$P(\alpha,\alpha^\ast)$ one can find a polynomial $Q(\alpha)$ of higher
degree in $\alpha$ (say $K$) such that the roots of $P$ are also roots
of $Q$. Since we know that $Q$ has at most $K$ solutions, it means that
$P$ can have at most $K$ solutions as well. For example, if we have
$P(\alpha,\alpha^\ast)=\alpha^2-\alpha^\ast=0$, we can complex conjugate
this equation and obtain
$P(\alpha,\alpha^\ast)^\ast=(\alpha^\ast)^2-\alpha=0$, which, upon
substituting $\alpha^\ast=\alpha^2$ we obtain
$Q(\alpha)=\alpha^4-\alpha=0$. This equation has four solutions
$(0,1,e^{i2\pi/3},e^{i2\pi/3})$ which are exactly the roots of $P$. Note
that we have been able to reduce the original equation to one containing
$\alpha$ only given the fact that the $P^\ast\ne P$. It is clear that
the polynomials fulfilling this condition are dense in the ring of all
polynomials, and this is the reason why we say that in the generic case
one can find $Q(\alpha)$.

In summary, in this Section we have shown that if there is a product
vector in $K(\rho)$ then we can find another operator $\rho_s$ supported
on $\C^2\otimes \C^{N-1}$ with the same separability properties as
$\rho$. In order to check whether we can have such product vectors, we
have shown that if the dimension of a subspace [e.g., $K(\rho)$] is
greater or equal to $N$, then we can always find product vectors in it.
On the other hand, if there is a product vector $|e,f\rangle \in
R(\rho)$ with $|e^\ast,f\rangle \in R(\rho^{T_A})$, then one can
decrease the rank of $\rho$, $\rho^{T_A}$ or both. This is always
possible if the ranks of $\rho$ and $\rho^{T_A}$ are sufficiently large.
Otherwise, one has to find roots of polynomials. We have also shown that
if there is a finite, sufficiently small number of product vectors
fulfilling this condition, then we automatically have a necessary and
sufficient condition for separability.

%=========================================================
\section{Results}

This section contains the main results of the paper concerning
separability of a density operator supported on $\C^2\otimes\C^N$. We
have divided them in four subsections. The first three deal with
different situations depending on the rank of $\rho$. Actually,
we just have to consider the cases in which $r(\rho)\ge N$ as
the following lemma states:

{\bf Lemma 12:} If $\rho$ is supported on $\C^2\otimes\C^N$
then $r(\rho)\ge N$.

{\em Proof:} Let us assume that $r(\rho)<N$. Then $K(\rho)$ has
dimension larger than $N$, and according to Lemma 10 it contains
a product vector. Using Lemma 7 we can find an operator
$\tilde\rho_1$ supported on $\C^2\otimes\C^{N-1}$ and with
$r(\tilde\rho_1)<N-1$. Again, $K(\tilde\rho_1)$ has dimension
larger than $N-1$, contains a product vector, and we can find an
operator $\tilde\rho_2$ supported on $\C^2\otimes\C^{N-2}$ and
with $r(\tilde\rho_2)<N-2$. We can proceed in the same way until
we have an operator $\tilde\rho_{N-1}$ supported on
$\C^2\otimes\C^{1}$ and with $r(\tilde\rho_2)<1$, which is
impossible.$\Box$

The first subsection shows that if $r(\rho)= N$ then $\rho$ is
separable. From that follows that bound entangled states have rank
larger than $N$. The second subsection deals with operators $\rho$ such
that $r(\rho)+r(\rho^{T_A})\le 3N$. It is shown that in the generic case
and for $N\le 10$ one can perform a constructive separability check. The
third subsection shows that given a density operator, one can always
subtract vectors until the condition $r(\rho)+r(\rho^{T_A})\le 3N$ is
obtained. In the fourth one, we show that if $\rho=\rho^{T_A}$ then
$\rho$ is separable. We also show that if $\rho$ and $\rho^{T_A}$ are
``sufficiently close" then $\rho$ is also separable.

%=========================================================
\subsection{rank$(\rho)=N$}

{\bf Theorem 1:} If $\rho$ is supported on $\C^2\otimes\C^N$
and $r(\rho)=N$ then $\rho$ is separable.

{\em Proof:} We use induction. For $N=1$ the statement is true. Let us
assume that it is true for $N-1$. We take now $\rho$ supported on
$\C^2\otimes \C^{N}$, with $r(\rho)=N$. Then $K(\rho)$ has dimension
$N$, and according to Lemma 10 it contains a product vector
$|e,f\rangle$. Using Lemma 7 we have that $\rho$ is separable iff
$\tilde\rho$ is separable. But $\tilde\rho$ is supported on $\C^2\otimes
\C^{N-1}$ and has rank $N-1$, and therefore it is separable by
assumption. $\Box$

{\bf Corollary 3a:} If $\rho$ is supported on $\C^2\otimes\C^N$
and $r(\rho)=N$, then $\rho$ can be written as a convex sum of
projectors on $N$ product vectors.

{\em Proof:} Just given by the construction of the Proof of Theorem 1.

{\bf Corollary 3b:} If $\rho$ is supported on $\C^2\otimes\C^N$
and $r(\rho)=N$ then $r(\rho^{T_A})=N$.

{\em Proof:} This is a direct application of the previous corollary
and Lemma 12.

{\bf Corollary 3c:} If $\rho$ is supported on $\C^2\otimes\C^N$
and is not separable, then $r(\rho)>N$.

{\em Proof:} Direct application of Lemma 12 and Theorem 1.

%=========================================================
\subsection{rank$(\rho)+$rank$(\rho^{T_A})\le 3N$}

We assume that $r(\rho),r(\rho^{T_A})>N$. We can use Lemma 11
with $H_1=R(\rho)$ and $H_2=R(\rho^{T_A})$ to determine the
conditions under which there exist product vectors that can be
subtracted from $\rho$. The problem reduces to finding the roots
of a polynomial in $\alpha$ and $\alpha^\ast$. As explained in
Section III, in the generic case these roots are also roots of
another polynomial in $\alpha$. If the corresponding projectors
$P_i$ are linearly independent [which imposes that the number of
such product vectors be smaller than the minimum of $r(\rho)^2)$
and $r(\rho^{T_A})^2$], then Lemma 11 provides us with a
necessary and sufficient condition for separability. We will
illustrate this with the case $\C^2\otimes\C^4$ in the following
section.

Let us now show that if $N\le 10$ then for a generic $\rho$ we
can always find less than the minimum of $r(\rho)^2$ and
$r(\rho^{T_A})^2$ product vectors $|e,f\rangle\in R(\rho)$ and
with $|e^\ast,f\rangle\in R(\rho)$. Without loss of generality
we can take $k(\rho)=X\ge k(\rho^{T_A})=Y$ (otherwise we can
substitute $\alpha\rightarrow\alpha^\ast$ in what follows). We
also assume that there are no product vectors in $K(\rho)$ since
otherwise we can trivially reduce the problem to
$\C^2\otimes\C^{N-1}$ decreasing the ranks of $\rho$ and
$\rho^{T_A}$ via Lemma 7. We therefore have $N> Y\ge X>0$ and
$X+Y\ge N$. Let us analyze separately the cases: (i) $X+Y=N$
and; (ii) $X+Y>N$. If $X+Y=N$ we have $0<X\le [N/2]$, where
$[W]$ denotes the integer part of $W$. According of Lemma 11,
the problem reduces to finding roots of a polynomial of degree
$X$ in $\alpha$ and $Y$ in $\alpha^\ast$. In Appendix A we give
a procedure to find those roots, which states that the number of
roots (for a generic case) will not exceed
$2^{X-1}[N^2+(N-X)^2-X(N-X)]$. This number should be smaller or
equal to $(N+X)^2$. One can check that for $N\le 10$ this
condition is always fulfilled. (ii) If $X+Y>N$ then we have
$[(N+1)/2]\le Y<N$. According to Lemma 11 there will be two (or
more) polynomial equations of degree $N-Y$ in $\alpha$ and $Y$
in $\alpha^\ast$. In Appendix B we show that in that case the
number of roots cannot exceed $2^{N-Y}Y$. This number should be
smaller or equal to $(2N-Y)^2$, which is always the case for
$N\le 10$.

Finally, using the results of the previous subsection we can
ensure that if $r(\rho)=r(\rho^{T_A})=N+1$ then either the
polynomials of Lemma 11 have exactly $N+1$ solutions, in which
case $\rho$ is separable and expresable in terms of $N+1$
product vectors, or it has none and then $\rho$ is bound
entangled. This is so because if one can subtract one product
vector, then one recovers the situation of Corollary 3a.

%=========================================================
\subsection{rank$(\rho)+$rank$(\rho^{T_A})> 3N$}

We can use Lemma 11 with $H_1=R(\rho)$ and $H_2=R(\rho^{T_A})$ to show
that there always exist product vectors that can be subtracted from
$\rho$. We can do that until we reach $r(\rho)+r(\rho^{T_A})\le 3N$. In
that case, we can use the criterion given in the previous subsection to
find out if the remaining operator is separable. Thus, this procedure
gives us a constructive sufficient criterion for separability.

%=========================================================
\subsection{Invariance under partial transposition: $\rho=\rho^T$}

{\bf Theorem 2:} If $\rho=\rho^{T_A}$ then $\rho$ is separable.

{\em Proof:} We prove it by induction. First, if $\rho$ is
supported on $\C^2\otimes\C^1$ then it is obviously true. Now,
let us assume that it is valid if $\rho$ is supported on
$\C^2\otimes\C^{N-1}$ and let us prove that then it is also
valid if it is on $\C^2\otimes\C^{N}$. According to Lemma 12, we
just have to consider two cases: (i) $r(\rho)= N$: Using Theorem
1 we have that $\rho$ is separable. (ii) $r(\rho)>N$: according
to Corollary 2 there is a product vector $|e_{\rm r},g\rangle$
with $|e_{\rm r}\rangle=|e_{\rm r}^\ast\rangle$. Using Corollary
1 we can use this product vector to reduce the rank of $\rho$.
But since $|e_{\rm r},g\rangle =|e_{\rm r}^\ast,g\rangle$ we
have that the resulting operator is also equal to its partial
transpose. We can proceed in this way until $r(\rho)=N$, in
which case we use Theorem 1 to show that $\rho$ is
separable.$\Box$

Note that we could have chosen a different basis for the partial
transposition. Moreover, since the property of separability is
independent of invertible local transformations $A$ acting on $\C^2$, we
have the following

{\bf Corollary 4:} If $[(A\otimes \one) \rho (A\otimes
\one)^\dagger]^{T_A}=(A\otimes \one) \rho (A\otimes \one)^\dagger$, for
some non singular operator $A$, then $\rho$ is separable.

Theorem 2 suggests that if $\rho$ is not very different from
$\rho^{T_2}$, then it should also be separable. Indeed, one can
construct a powerful sufficient separability condition based on that
theorem. For that we have to introduce some definitions. In $2\times N$
we can always write $\rho$ as
\be
\label{dec2}
\rho=\frac{\rho+\rho^{T_A}}{2} + \frac{\rho-\rho^{T_A}}{2}
\equiv \rho_s + \sigma^A_y\otimes B,
\ee

Here $2\rho_s=\rho+\rho^{T_A}$, $\sigma^A_y=i(|0\rangle_A\langle
1|-|1\rangle_A\langle 0|)$, and $4B=4B^\dagger={\rm
tr_A}[\sigma_y^A(\rho-\rho^{T_A})]$. This operator $B$ can be
decomposed as
\be
B=\sum_{i=1}^K \lambda_i |v_i\rangle\langle v_i|.
\ee
In particular, one of such is the spectral decomposition. Given one of
such decomposition $\{\lambda_i,|v_i\rangle\}_{i=1}^K$ and a set of real
numbers $\{a_i\}_{i=1}^K$ we define the operator
\be
C(a,\lambda,v)\equiv \sum_{i=1}^K |\lambda_i|
(a_i^2|0\rangle\langle 0| +a_i^{-2}|1\rangle\langle 1|)\otimes
|v_i\rangle\langle v_i|,
\ee
which is obviously positive. We have:

{\bf Theorem 3:} Given a decomposition of $B$
$\{\lambda_i,|v_i\rangle\}_{i=1}^K$ and a set of real numbers
$\{a_i\}_{i=1}^K$, if $\|C^{1/2}(a,\lambda,v)\rho_s^{-1/2}\|^2\le 1$,
then $\rho$ is separable.

{\em Proof:} We define $\tilde\rho_s=\rho_s-C(a,\lambda,v)=
\tilde\rho_s^{T_A}\ge 0$ according to Lemma 1. Using Theorem 2 we have
that $\tilde\rho_s$ is separable. Let $|w_i\rangle=a_i|0\rangle
-ia_i^{-1} {\rm sign}(\lambda_i)|1\rangle$. Then, it is easy to check
that
\be
\label{end}
\rho=\tilde\rho_s + \sum_{i=1}^K|\lambda_i| |w_i,v_i\rangle\langle w_i,v_i|.
\ee
which shows that $\rho$ is separable.$\Box$

Thus, we can show that a density operator is separable if we can find a
decomposition of $B$ and a set of real numbers that fulfill certain
conditions. In particular we can take the spectral decomposition of $B$
and $a_i=1$. Using the fact that $\|AB\| \le \|A\| \|B\|$ one can easily
prove the following:

{\bf Corollary 5:} If $\rho+\rho^{T_A}$ is of full range and
$\|(\rho +
\rho^{T_A})^{-1}\| \|\rho-\rho^{T_A}\| \le 1$, then $\rho$ is separable.
This corollary implies that if $\rho$ is full range and is very close to
$\rho^{T_A}$ then it is separable.

Finally, let us mention that we have taken here all partial transposes
with respect to the first system. One may wonder whether if $\rho$ is
invariant under partial transposition with respect to the second system,
then Theorem 2 still holds. Actually, this is not the case. A
counterexample can be found in Ref.\ \cite{Ho98b}, by simply noting that
the bound entangled state given there is equal to its partial transpose
if we take that in a different basis.

% ===============================================================
\section{Example: $\C^2\otimes\C^4$}

As an illustration of the methods developed in the previous sections to
deal with separability when one has low rank density operators, here we
will show how one can apply them in the case of $\C^2\otimes \C^4$. We
will assume that: (i) $\rho$ has no product vector in its kernel since
otherwise using Lemma 7 we can reduce the problem to $\C^2\otimes \C^3$
and there we know that $\rho$ is separable; (ii)
$r(\rho),r(\rho^{T_A})>4$ since otherwise using Theorem 1 we know that
$\rho$ is separable; (iii) $r(\rho)+r(\rho^{T_A})\leq 12$ since in that
case we can use our separability criterion. This section is divided into
$3$ parts:

\begin{itemize}
\item $r(\rho)=r(\rho^{T_{A}})=5$
\item $r(\rho)+r(\rho^{T_{A}})\leq 12$
and $r(\rho)\not=r(\rho^{T_{A}})$.
\item $r(\rho)=r(\rho^{T_{A}})=6$
\end{itemize}

As before, we will denote by $\{|\Psi_{i}^{1}\rangle, i=1\ldots
k(\rho)\}$ and $\{|\Psi_{i}^{2}\rangle, i=1\ldots
k(\rho^{T_{A}})\}$ a basis in $K(\rho)$ and $K(\rho^{T_A})$,
respectively. We will use the fact that, according to Lemma 11,
there exists a product vector $|e,f\rangle\in R(\rho)$ and
$|e^\ast,f\rangle \in R(\rho^{T_A})$ iff we can find an $\alpha$
such that there are at most $3$ linearly independent vectors
among $\{\alpha \langle \Psi_{i_1}^{1}|0\rangle +
\langle \Psi_{i_1}^{1}|1\rangle, \alpha^\ast
\langle \Psi_{i_2}^{2}|0\rangle + \langle \Psi_{i_2}^{2}|1\rangle\}$.

%========================================================
\subsection{rank($\rho$)=rank($\rho^{T_{A}}$)$=5$}

Each of the two kernels has dimension $3$ and therefore, according to
Lemma 11, in order to find subtractable product vectors we have to solve
$M(\alpha,\alpha^\ast)\vec{f}=0$, where
\be
M(\alpha,\alpha^\ast)= \left( \begin{array}{c}
 \alpha
\langle \Psi_{1}^{1}|0\rangle + \langle
\Psi_{1}^{1}|1\rangle\\
 \alpha
\langle \Psi_{2}^{1}|0\rangle + \langle
\Psi_{2}^{1}|1\rangle\\
 \alpha
\langle \Psi_{3}^{1}|0\rangle + \langle
\Psi_{3}^{1}|1\rangle\\
 \alpha^\ast \langle \Psi_{1}^{2}|0\rangle
+ \langle \Psi_{1}^{2}|1\rangle\\
 \alpha^\ast \langle \Psi_{2}^{2}|0\rangle
+ \langle \Psi_{2}^{2}|1\rangle\\
 \alpha^\ast \langle \Psi_{3}^{2}|0\rangle
+ \langle \Psi_{3}^{2}|1\rangle
\end{array}\right),
 \ee
is a $6\times 4$ matrix. Using Lemma 9 we know that the last three rows
in $M$ are linearly independent. Thus, this equation has non trivial
solutions iff
\bma
\bea
\mbox{det}_{1}(\alpha,\alpha^\ast)=\mbox{det}[M_{1}(\alpha,\alpha^\ast)]&=&0\\
\mbox{det}_{2}(\alpha,\alpha^\ast)=\mbox{det}[M_{2}(\alpha,\alpha^\ast)]&=&0\\
\mbox{det}_{3}(\alpha,\alpha^\ast)=\mbox{det}[M_{3}(\alpha,\alpha^\ast)]&=&0,
\eea
\ema
where $M_{1,2,3}$ are $4\times 4$ matrices formed by the last three rows
of $M$ and the $1$--st, $2$--nd, and $3$--rd rows, respectively. Note
that $\mbox{det}_{1}$, $\mbox{det}_{2}$ and $\mbox{det}_{3}$ are
polynomials linear in $\alpha$ and cubic in $\alpha^\ast$. Thus, we can
write them as
\bma
\label{eee}
\bea
\label{eq:det1}
\mbox{det}_{1}&=&\alpha P^{1}_{3}(\alpha^\ast) +P^{0}_{3}(\alpha^\ast)=0\\
\label{eq:det2}
\mbox{det}_{2}&=&\alpha \tilde P^{1}_{3}(\alpha^\ast) +\tilde P^{0}_{3}(\alpha^\ast)=0\\
\label{eq:det3}
\mbox{det}_{3}&=&\alpha \hat P^{1}_{3}(\alpha^\ast) +\hat P^{0}_{3}(\alpha^\ast)=0,
\eea
\ema
where the subscripts denote the degree of the polynomials. In
order to solve this equation we treat $\alpha$ and $\alpha^\ast$
as two independent variables. Multiplying Eq.\ (\ref{eq:det1})
by $\tilde P^{1}_{3}(\alpha^\ast)$, Eq.\ (\ref{eq:det2}) by
$P^{1}_{3}(\alpha^\ast)$ and subtracting them we find an
polynomial equation for $\alpha^\ast$ of degree $6$. Doing the
same thing but with Eq.\ (\ref{eq:det3}) instead of Eq.\
(\ref{eq:det2}) gives another polynomial equation for
$\alpha^\ast$ of 6th degree. Subtracting these two polynomials
we obtain a polynomial of degree 5. Thus, there are at most 5
solutions (i.e. 5 product vectors, as stated in Section IV B).
Note that the solutions have still to fulfill all Eq.\
(\ref{eee}). Note also that if we know $\alpha$ we can calculate
the kernel of the matrix $A$ for this particular value and find
so the corresponding $|f\rangle$.

% ========================================================
\subsection{r($\rho$)+r($\rho^{T_{A}}$)$\leq 12$ where
r($\rho$) $\neq$ r($\rho^{T_{A}}$)}

We are going to divide this part itself into:
\begin{itemize}
\item r($\rho$)$=6$
and r($\rho^{T_{A}}$)$=5$ or r($\rho$)$=5$ and
r($\rho^{T_{A}}$)$=6$,
\item r($\rho$)$=7$
and r($\rho^{T_{A}}$)$=5$ or r($\rho$)$=5$ and
r($\rho^{T_{A}}$)$=7$.
\end{itemize}
Let us begin with the case where r($\rho$)$=6$ and
r($\rho^{T_{A}}$)$=5$. Then the dimension of the kernel of $\rho$ is $2$
and the one of the kernel of $\rho^{T_{A}}$ is $3$. Thus we can write
the equations in the same way as in the previous case, but there are
only two of them. We have
\bma
\bea
\mbox{det}_{1}=\mbox{det}[M_{1}(\alpha,\alpha^\ast)]&=&0\\
\mbox{det}_{2}=\mbox{det}[M_{2}(\alpha,\alpha^\ast)]&=&0,
\eea
\ema
where $M_{1}$ and $M_{2}$ are the same matrices as above. Thus
the way of calculating the $\alpha$'s is exactly the same as
before, obtaining at most 6 solutions for $\alpha^\ast$. On the
other hand it is clear that the same can be done in the case
where r($\rho$)=$5$ and r($\rho^{T_{A}}$)$=6$.

Now we consider the case r($\rho$)$=7$ and r($\rho^{T_{A}}$)$=5$. Here
we have only one equation that has to be fulfilled. We find
\be
\mbox{det}_{1}=\mbox{det}[M_{1}(\alpha,\alpha^\ast)]=0,
\ee
which leads to
\bma
\bea
\label{75}
0&=&\alpha P^{1}_{3}(\alpha^\ast)+P^{0}_{3}(\alpha^\ast)\\
 &=&(\alpha^\ast)^3 Q^{3}_{1}(\alpha)+(\alpha^\ast)^2
Q^{2}_{1}(\alpha)+\alpha^\ast Q^{1}_{1}(\alpha)+
Q^{0}_{1}(\alpha)
\label{76}
\eea
\ema
Taking the complex conjugate of Eq.\ (\ref{75}) we have:
\be
\alpha^\ast \bar P^{1}_{3} (\alpha)+ \bar P^{0}_{3}(\alpha)=0,
\label{752}
\ee
where $\bar P$ is the same polynomial as $P$ but with the
coefficients complex conjugated. As explained in the Appendix we
are using now Eq.\ (\ref{752}) to transform the polynomial in
Eq.\ (\ref{75}) in one depending only on $\alpha$. That is, we
multiply Eq.\ (\ref{76}) by $\bar P^{1}_{3} (\alpha)$ and Eq.\
(\ref{752}) by $ Q^{3}_{1} (\alpha)$ and subtract them. We do
the same two more times and end up with a polynomial of 10--th
degree in $\alpha$. For the different values of $\alpha$ we can
now calculate the kernel of the matrix $M$ and obtain
$|f\rangle$. Thus we have all the product vectors $|e,f\rangle$
fulfilling $|e,f\rangle \in R(\rho)$ and $|e^\ast,f\rangle \in
R(\rho^{T_A})$.

% ======================================================
\subsection{rank($\rho$)=rank($\rho^{T_{A}}$)$=6$}

Here we have two vectors in both, the kernel of $\rho$ and the one of
$\rho^{T_A}$. Thus the equation that has to be fulfilled is
\bma
\bea
\label{eq:dett}
0=\mbox{det}(M(\alpha,\alpha^\ast))&=& \alpha^2
P^{2}_{2}(\alpha^\ast)+
\alpha P^{1}_{2}(\alpha^\ast)+P^{0}_{2}(\alpha^\ast)\\
&=&(\alpha^\ast)^2 {Q}^{2}_{2}(\alpha)+\alpha^\ast
Q^{1}_{2}(\alpha)+Q^{0}_{2}(\alpha).
\label{eq:dett1}
\eea
\ema
Again, we take the complex conjugate of Eq.\ (\ref{eq:dett}) and
deal with the following two equations:
\bma
\bea
\label{eq:first}
(\alpha^\ast)^2 Q^{2}_{2}(\alpha)+\alpha^\ast
Q^{1}_{2}(\alpha)+Q^{0}_{2}(\alpha)&=&0\\
\label{eq:second}
(\alpha^\ast)^2 \bar{Q}^{2}_{2}(\alpha)+\alpha^\ast
\bar{Q}^{1}_{2}(\alpha)+\bar{Q}^{0}_{2}(\alpha)&=&0.
\eea
\ema
First, we multiply Eq.\ (\ref{eq:first}) by $\bar{Q}^{2}_{2}(\alpha)$,
Eq.\ (\ref{eq:second}) by $Q^{2}_{2}(\alpha)$ and subtract them. Then we
multiply Eq.\ (\ref{eq:first}) by $\bar{Q}^{0}_{2}(\alpha)$, Eq.\
(\ref{eq:second}) by $Q^{0}_{2}(\alpha)$ and subtract them and divide
the resulting polynomial by $\alpha^\ast$. Thus we find
\bma
\bea
\label{eq:i}
&\alpha^\ast& [Q^{1}_{2}(\alpha)\bar{Q}^{2}_{2}(\alpha)
-Q^{2}_{2}(\alpha)\bar Q^{1}_{2}(\alpha)]+ \nonumber\\
&&Q^{0}_{2}(\alpha) \bar{Q}^{2}_{2}(\alpha)
-\bar{Q}^{0}_{2}(\alpha)Q^{2}_{2}(\alpha)=0\\ &\alpha^\ast&
[Q^{2}_{2}(\alpha) \bar{Q}^{0}_{2}(\alpha)
 -\bar{Q}^{2}_{2}(\alpha)Q^{0}_{2}(\alpha)]+ \nonumber\\
 && Q^{1}_{2}(\alpha)\bar{Q}^{0}_{2}(\alpha)-Q^{0}_{2}(\alpha)
   \bar Q^{1}_{2}(\alpha)=0.
\label{eq:j}
\eea
\ema
Now we multiply Eq.\ (\ref{eq:i}) by $[Q^{2}_{2}(\alpha)
\bar{Q}^{0}_{2}(\alpha) -\bar{Q}^{2}_{2}(\alpha)Q^{0}_{2}(\alpha)]$,
Eq.\ (\ref{eq:j}) by
$[Q^{1}_{2}(\alpha)\bar{Q}^{2}_{2}(\alpha)-Q^{2}_{2}(\alpha)\bar
Q^{1}_{2}(\alpha)]$ and subtract them again and end up with a polynomial
of degree $8$ in $\alpha$. Note that this means that it may
not be possible to write the density operator as a convex sum
of 6 product vectors. In fact, we have confirmed that and found
examples in which one needs 8 product vectors.

%===============================================================
\appendix

\section*{Reduction of one polynomial}

In this appendix we show that given a generic polynomial
$P_{X,Y}(\alpha,\alpha^\ast)$ of degree $X$ and $Y$ in $\alpha$ and
$\alpha^\ast$ respectively, there are at most $2^{X-1}[X+Y(Y-X+1)]$
roots, where without loss of generality we have assumed $Y\ge X$. The
idea is to find another polynomial $Q(\alpha)$ of degree
$2^{X-1}[X+Y(Y-X+1)]$ such that all the roots of $P$ are also roots of
$Q$. Since $Q$ has at most $2^{X-1}[X+Y(Y-X+1)]$ roots, this will prove
our statement.

We start by writing the equation $P=0$ as
\bma
\bea
\label{Eq1}
0=P_{X,Y}(\alpha,\alpha^\ast) &=& \sum_{k=0}^X \alpha^k
P_Y^k(\alpha^\ast)\\ &=& \sum_{k=0}^Y (\alpha^\ast)^k
Q_X^k(\alpha),
\eea
\ema
where $P_Y^k$ and $Q_X^k$ denote polynomials of degree $Y$ and
$X$, respectively. We complex conjugate the last line in this
equation and obtain
\be
\label{Eq2}
0= \sum_{k=0}^Y \alpha^k \bar Q_X^k(\alpha^\ast),
\ee
where $\bar Q$ is the same polynomial as $Q$ but with the
coefficients complex conjugated. From now on we treat $\alpha$
and $\beta\equiv\alpha^\ast$ as two independent variables. Thus,
we can write Eqs.\ (\ref{Eq1}) and (\ref{Eq2}) as
\bma
\bea
\label{X}
\sum_{k=0}^X \alpha^k P_Y^k(\beta)&=&0\\
\label{Y}
\sum_{k=0}^Y \alpha^k \bar Q_X^k(\beta)&=&0.
\eea
\ema

We proceed now in two steps:

(1) If $Y\ne X$ we use Eq.\ (\ref{X}) to transform Eq.\
(\ref{Y}) to one of the form
\be
\label{Z}
\sum_{k=0}^X \alpha^k R_Z^k(\beta)=0,
\ee
where $Z=X+Y(Y-X)$. This can be done in the following way. First
we multiply Eq.\ (\ref{X}) by $\alpha^{Y-X}\bar Q_X^Y(\beta)$,
Eq.\ (\ref{Y}) by $P_Y^X(\beta)$, and subtract them. In this way
we obtain a polynomial equation of degree $Y-1$ in $\alpha$ and
$Y+X$ in $\beta$. We use again Eq.\ (\ref{X}) to transform this
one to a polynomial equation of degree $Y-2$ in $\alpha$. We
proceed in the same way until we reach Eq.\ (\ref{Z}). Note that
if $Y=X$ then we have automatically Eq.\ (\ref{Z}) with $Z=X$.

(2) First, we multiply Eq.\ (\ref{X}) by $R_Z^X(\beta)$, Eq.\
(\ref{Z}) by $P_Y^X(\beta)$ and subtract them. In parallel, we
multiply Eq.\ (\ref{X}) by $R_Z^0(\beta)$, Eq.\ (\ref{Z}) by
$P_Y^0(\beta)$, subtract them, and divide by $\alpha$ the
result. In this way, we obtain two polynomial equations of
degree $X-1$ in $\alpha$ and $Y+Z$ in $\beta$. We can do the
same thing with these two, and proceed in the same vein until we
reach a polynomial equation of the form $\bar Q(\beta)=0$ (i.e.
$Q(\alpha)=0$) of degree $2^{(X-1)}(Z+Y)$. All the roots of the
original polynomial must be roots of $Q$.

\section{Reduction of two polynomials}

Following a similar procedure as in the previous appendix, we
show here that given two generic polynomials $P_{X,Y}$ and
$\tilde P_{X,Y})$ of degree $X$ and $Y$ in $\alpha$ and
$\alpha^\ast$ respectively, there are at most $2^{X}Y$ common
roots, where without loss of generality we have assumed $Y\ge
X$. Treating $\alpha$ and $\beta\equiv\alpha^\ast$ as
independent variables, if we write
\bma
\bea
0 &=& P_{X,Y}(\alpha,\alpha^\ast) = \sum_{k=0}^X \alpha^k
P_Y^k(\beta),\\ 0 &=& \tilde P_{X,Y}(\alpha,\alpha^\ast) =
\sum_{k=0}^X \alpha^k \tilde P_Y^k(\beta),\\
\eea
\ema
we can use the same procedure presented in step (2) in the
previous appendix to obtain a polynomial equation $Q(\alpha)$ of
degree $2^X Y$.

We emphasize that if one has more polynomials, one can further
decrease the degree of $Q(\alpha)$. An example is shown in Section V,
when the ranks of $\rho$ and $\rho^{T_A}$ equal 5.

% -------------------------------------------------------------

This work has been supported by Deutsche Forshungsgemeinschaft
(SFB 407) and Schwerpunkt "Quanteninformationsverarbeitung", the
\"Osterreichisher Fonds zur F\"rderung der wissenschaftlichen
Forschung (SFB P11), the European TMR network
ERB-FMRX-CT96-0087, and the Institute for Quantum Information
Gmbh. J. I. C. thanks the University of Hannover for
hospitality. We thank, P. Horodecki, A. Sanpera and G. Vidal for
discussions.

% -------------------------------------------------------------

\end{document}